\documentclass[preprint,showpacs,preprintnumbers,amsmath,amssymb]{revtex4}

\usepackage{graphicx}% Include figure files
\usepackage{dcolumn}% Align table columns on decimal point
\usepackage{bm}% bold math

\input epsf.tex
\def\DESepsf(#1 width #2){\epsfxsize=#2 \epsfbox{#1}}
\begin{document}

%\preprint{APS/123-QED}

\title{Extraction of Axial Form Factors from
$\overline B^0\to D^+\,K^-\,K^{*0}$ Decay}
% Force line breaks with \\
\vspace{10cm}

\author{$^1$Chun-Khiang Chua, $^2$Wei-Shu Hou and $^2$Shang-Yuu Tsai}
\affiliation{$^1$Institute of Physics, Academia Sinica, Taipei,
Taiwan 115, Republic of China\\
%\author{}
%\affiliation{%
$^2$Department of Physics, National Taiwan University, Taipei,
Taiwan 106, Republic of China
}%

%% \homepage{http://www.Second.institution.edu/~Charlie.Author}

\date{\today}% It is always \today, today,

  %Bottom mesons

\begin{abstract}
Based on $\overline B\to D^{(*)}K^-K^{*0}$ decay data, which
supports factorization, we propose to extract the kaon axial form
factor in the factorization framework. Experiment indicates that
the $K^-K^{*0}$ pair is produced by an axial current where only
one out of three axial form factors is dominant. %, with a significant contribution from $a_1(1260)$ resonance.
The axial form factor can be extracted by fitting the $K^-K^{*0}$
mass spectrum with an $a_1(1260)$-resonance plus QCD-motivated
non-resonant contributions, which can be improved as data
improves.
\end{abstract}

\maketitle

%%%%%%%%%%%%%%%%%%%%%%
\section{Introduction}
\label{sec:Introduction}
%%%%%%%%%%%%%%%%%%%%%%

The three-body $\overline B\to DK^-K^{*0}$, $D^*K^-K^{*0}$ and
$\overline B\to DK^-K^0$, $D^*K^-K^0$ decays were observed for the
first time by the Belle Collaboration based on $29.4$~fb$^{-1}$ of
data~\cite{Drutskoy:2002ib}. For the $\overline B\to
D^{(*)}K^-K^{*0}$ channels, one has a peak near threshold in the
$K^-K^{*0}$ mass spectrum, which can be described by a dominant
$a_1(1260)$ resonance contribution, although there is a possible
peak at $\sim2$~GeV. In addition, angular analysis finds
$K^-K^{*0}$ to be in the $J^P=1^+$ configuration, supporting the
$a_1(1260)$ pole dominance picture. The fitted parameters,
however, give larger $a_1^-(1260)\to K^-K^{*0}$ branching
fraction~\cite{Drutskoy:2002ib}, by a factor of 2 or more, than
what is obtained from the three pion mass spectrum in $\tau$
decays~\cite{Asner:1999kj}.

The peaking near threshold of the $K^-K^{*0}$ mass spectrum
suggests a {\it quasi two-body} process where the colinear
$K^-K^{*0}$ pair recoils against the $D^{(*)}$ meson. This suggests that
factorization could be at work for such three-body decays. In our
previous work on $\overline B^0\to D^{(*)+}K^-K^{(*)0}$ and
$B^-\to D^{(*)0}K^-K^{(*)0}$ decays, two kinds of decay amplitudes
arose due to different flavor structures that give rise to the
$D^{(*)}$ meson~\cite{Chua:2002pi}. The $\overline B^0\to
D^{(*)+}K^-K^{(*)0}$ process involves only the matrix element
$\langle K^-K^{(*)0}|V-A|0\rangle$, where the $K^-K^{(*)0}$ is
produced by a weak $V-A$ current. For $B^-\to
D^{(*)0}K^-K^{(*)0}$, one has in addition a $\langle
K^-K^{(*)0}|V-A|B^-\rangle$ contribution, where $B^-$ goes into
$K^-K^{(*)0}$ via a weak current.

\begin{figure*}[t!]
\includegraphics[width=3.1in]{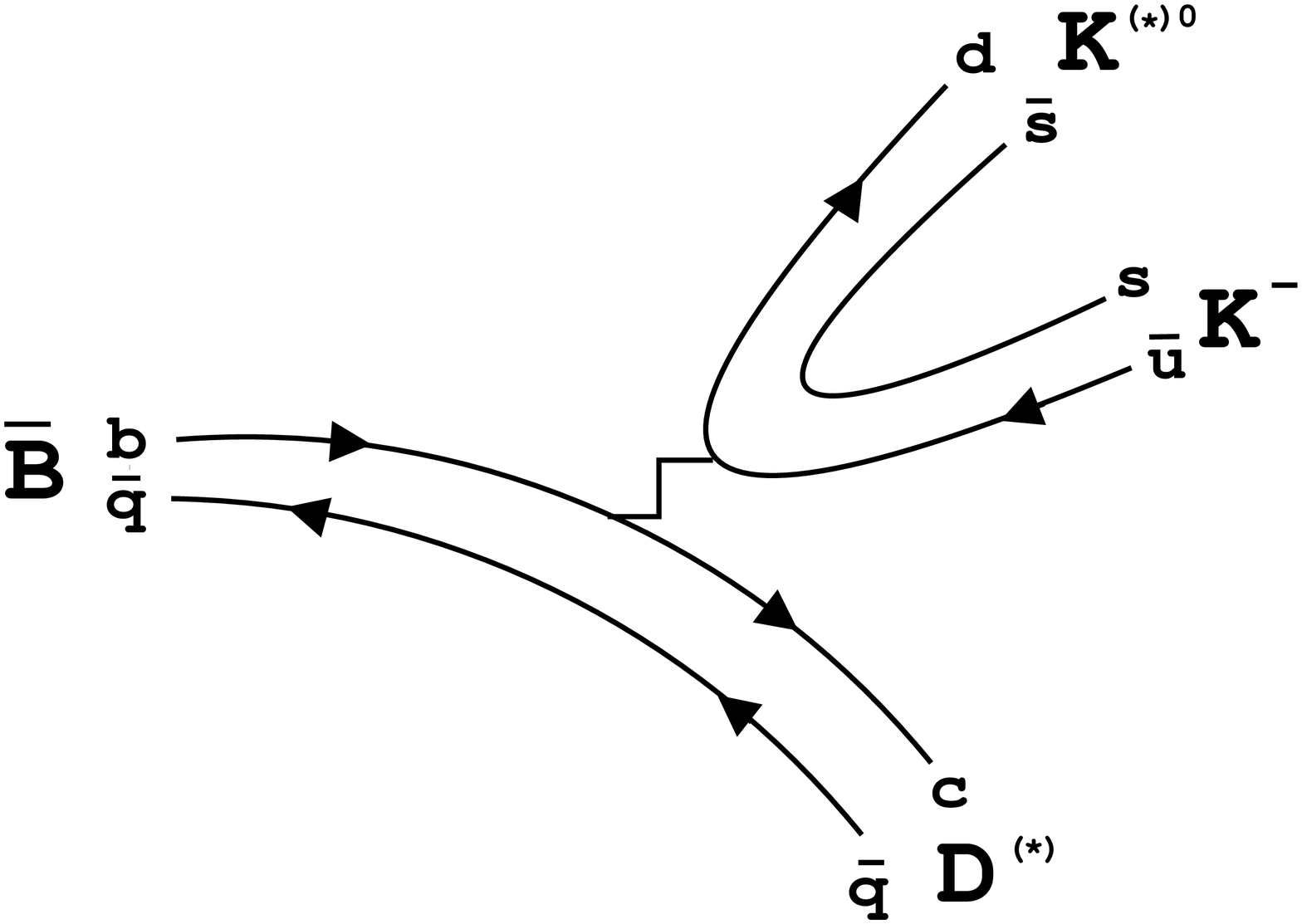}
\includegraphics[width=3.1in]{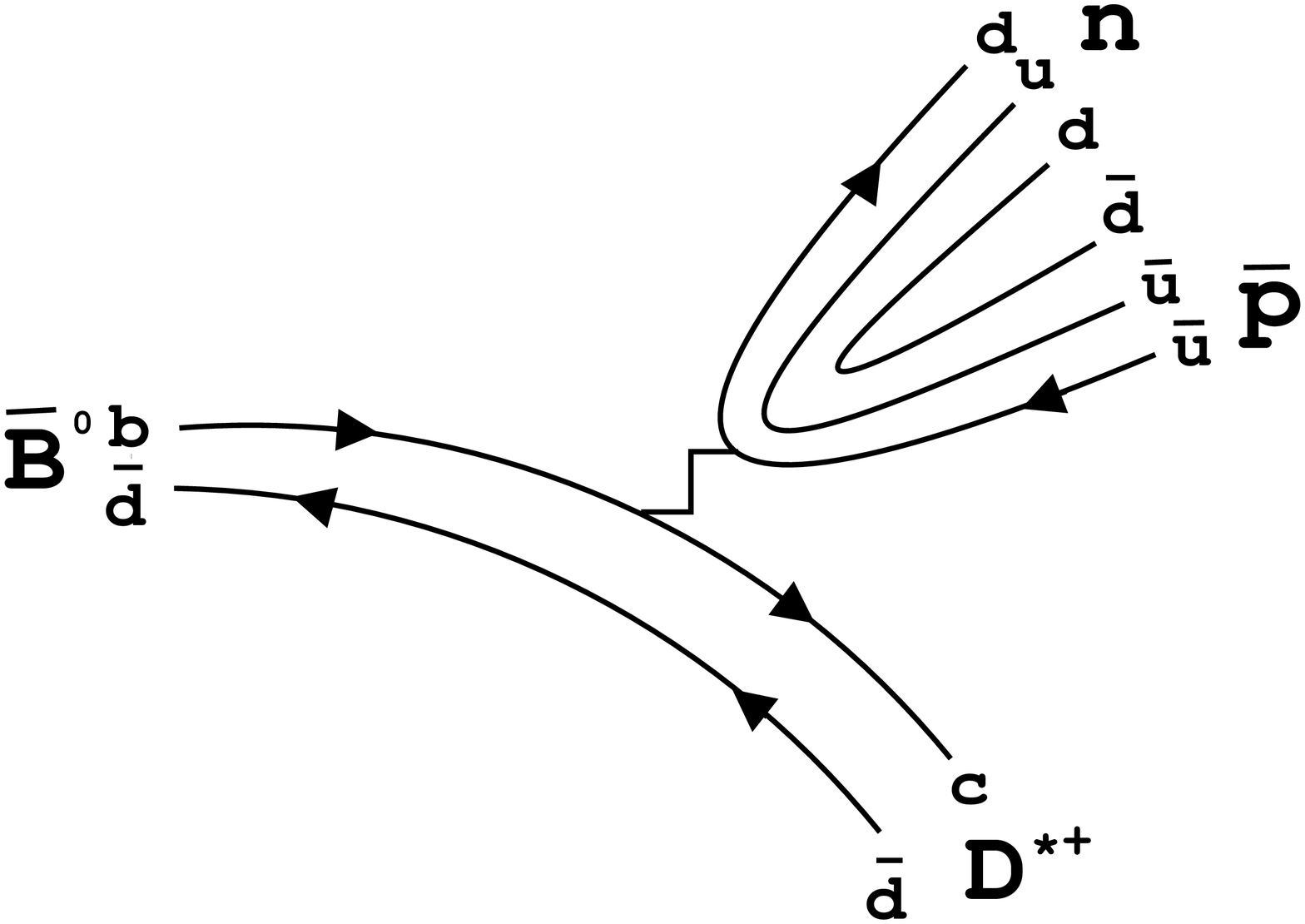}
\caption{\label{fig:current} (a) The $K^-K^{(*)0}$ and (b) $\bar p
n$ pairs produced by a current. The wavy line is the $W$ boson.}
\end{figure*}

Under factorization, the $\overline B^0\to D^{(*)+}K^-K^0$ decay
amplitude is a product~(Fig.~\ref{fig:current}(a)) of the matrix
elements $\langle K^-K^0|V-A|0\rangle$ and $\langle
D^{(*)+}|V-A|\overline B^0\rangle$. From parity, $K^-K^0$ can only
be produced by the vector current in $\langle
K^-K^0|V-A|0\rangle$, which carries both $J^P=0^+$ and $1^-$
components. The $K^-K^0$ pair, however, should dominantly be in
the $1^-$ state by isospin symmetry. This is consistent with
experiment, which finds $K^-K^0$ dominantly in
$1^-$~\cite{Drutskoy:2002ib,remark}. By using isospin rotation,
the kaon weak form factor $\langle K^-K^0|V|0\rangle$ can be
further related to the kaon electromagnetic~(em) form factors in
$e^+e^-$ annihilation, where much data exist. Without tuning
parameters, we obtained the value for $\mathcal{B}(\overline
B^0\to D^{(*)+}K^-K^0)$ which is in good agreement with
experiment~\cite{Chua:2002pi}. The predicted $K^-K^0$ mass
spectrum has a peak near threshold as a consequence of the kaon
form factor, which can be checked by experiment. The prediction of
$K^-K^0$ to be in the $1^-$ state and good agreement with the
observed branching fraction give strong support for factorization
in $\overline B^0\to D^{(*)+}K^-K^0$ decay.

%
%-----------------------------------------------------------------------
%------------------------------------------------------------------------
%

As illustrated in Fig.~\ref{fig:current}, $\overline B^0\to
D^{(*)+}K^-K^0$ decay bears some similarity with the baryonic
$B^0\to D^{*-}p\bar n$ decay, which was first measured by the CLEO
Collaboration~\cite{Anderson:2000tz}. Under factorization, the
$B^0\to D^{*-}p\bar n$ decay amplitude contains~\cite{Chua:2001vh}
the matrix element $\langle p\bar n|V-A|0\rangle$, which involves
both vector and axial current form factors. We had also used
isospin to relate the vector current form factor to the EM data.
It was found that the vector current contribution could account
for $\sim60\%$ of the observed rate~\cite{Chua:2001vh}. The
predicted $p\bar n$ mass spectrum also exhibits near threshold
enhancement. Unfortunately, this mass spectrum cannot be checked
immediately. Furthermore, there is no data on axial current baryon
form factors. We suggested instead that one could perform an
inverse transform to obtain nucleon axial form factors once the B
decay data becomes available.
%, since there is no interference between the vector and
% the axial current contributions.
It would be interesting if fundamental quantities such as nucleon
axial form factors can be extracted directly from $B$ physics
data.
It should be noted that $\overline B {}^0\to D^{(*)0}p\bar p$
spectra have been measured~\cite{Abe:2002tw} and the $a_1$-pole
contribution could be important~\cite{Cheng:2002fp}. However, we
cannot use them to obtain nucleon axial form factors, since these
modes are dominated by transition contributions, i.e. ${\mathcal
A}\propto\langle p\bar p|j|\overline B{}^0\rangle~\langle
D^{(*)0}|j|0\rangle$, in the factorization approach.

Unlike the $B^0\to D^{*-}p\bar n$ case, we already have some data
of the $K^-K^{*0}$ mass spectrum for the $\overline B\to
D^{(*)}K^-K^{*0}$ decay. In the factorization framework, the
$K^-K^{*0}$ pair in $\overline B^0\to D^{(*)+}K^-K^{*0}$ can be
produced by both vector and axial currents. However, experiment
found that $K^-K^{*0}$ is dominantly in $J^P=1^+$ configuration,
implying a dominant axial current contribution. By neglecting
contributions from the vector and the timelike component of the
axial current~(which would give $0^-$), it is possible to obtain
the timelike $K^-$-$K^{*0}$ axial form factors from $\overline
B^0\to D^{(*)+}K^-K^{*0}$ decay data.

In this paper, our purpose is to demonstrate the extraction of
axial form factor from data. Unfortunately, although more data is
already available, at present the $K^-K^{*0}$ mass spectrum is
given only by combining $D^+K^-K^{*0}$, $D^{*+}K^-K^{*0}$,
$D^0K^-K^{*0}$ and $D^{*0}K^-K^{*0}$ modes. To stimulate further
experimental studies, we generate mock data based on the existing
one, from which an inverse transform is done to obtain the axial
form factor. In the next section we formulate the factorization
approach and define the $K^-$-$K^{*0}$ axial form factors. The
implications of angular analysis on the form factors will be
examined, and the form factor is then parameterized for subsequent
use. The fit results to mock data are obtained in
Sec.~\ref{sec:Results}, which is followed by discussion and
conclusion in the last section.

%%%%%%%%%%%%%%%%%%%%%%
\section{Formulation}
\label{sec:Formulation}
%%%%%%%%%%%%%%%%%%%%%%

The relevant effective Hamiltonian is
\begin{equation}
{\mathcal H_{\rm eff}}= \frac{G_F}{\sqrt2} V_{cb} V_{ud}^* \big[
   c_1(\mu)\mathcal O^c_1(\mu)+c_2(\mu)\mathcal O^c_2(\mu)\big]\,,
\label{eq:Heff}
\end{equation}
where $c_i(\mu)$ are the Wilson coefficients and $V_{cb}$ and
$V_{ud}$ are the Cabibbo-Kobayashi-Maskawa~(CKM) matrix elements.
The operators ${\mathcal O}_i$ are products of $V-A$ currents,
i.e. ${\mathcal O^c_1}=(\bar c b)_{V-A}\,(\bar d u)_{V-A}$ and
${\mathcal O^c_2}=(\bar d b)_{V-A}\,(\bar c u)_{V-A}$, where
$(\bar qq^\prime)_{V-A}\equiv \bar
q\gamma^\mu(1-\gamma^5)q^\prime$.

We concentrate on the $\overline B^0\to D^+K^-K^{*0}$ decay mode,
since under factorization it involves only a simple $\overline
B^0\to D^+$ transition and $\langle K^-K^{*0}|V-A|0\rangle$,
giving the decay amplitude
\begin{equation}
{\mathcal A}(D^+K^-K^{*0})= \frac{G_F}{\sqrt2} V_{cb} V_{ud}^*
\,a_1 \langle D^+|(\bar c b)_{V-A}|\overline B {}^0\rangle
      \langle K^-K^{*0}|(\bar d u)_{V-A}|0\rangle\,,
\label{eq:amplitude}
\end{equation}
where $a_1\equiv c_1+c_2/N_c$ with $N_c$ the effective number of
colors if naive factorization is used. In practice, we may need to
use the $a_1$ coefficient fitted from $\overline B\to D a_1$
decays. The matrix element $\langle D^+|(\bar c b)_{V-A}|\overline
B {}^0\rangle$ is the same as in two-body decay, and is
parameterized by
\begin{equation}
\langle D^+(p_D)\left| V^\mu\right|\overline B^0(p_B)\rangle=
\bigg(g^{\mu\nu}-\frac{q^\mu q^\nu}{q^2}\bigg)(p_B+p_D)_\nu
F_1^{BD}(q^2)+\frac{m^2_B-m^2_D}{q^2}\,q^\mu F_0^{BD}(q^2)\,,
\label{eq:B2D}
\end{equation}
where $q\equiv p_B-p_D=p_K+p_{K^*}$. We employ the Melikhov-Stech
(MS) model~\cite{Melikhov:2000yu} for the form factors $F^{BD}_0$
and $F^{BD}_1$. For $\langle K^-K^{*0}|(\bar d u)_{V-A}|0\rangle$,
since $K^-K^{*0}$ is produced by the $V$ and $A$ currents, in the
$K^-K^{*0}$ rest frame, the allowed angular momentum and parity
configurations for $K^-K^{*0}$ are $J^P=0^-$, $1^+$ and $1^-$,
which transform in the same way as the axial vector current
$A^\mu$~($0^-, 1^+$) and the spatial part of the vector current
$V^i$~($1^-$), respectively. A convenient parametrization that
manifests the possible $J^P$ configurations is then given by
\begin{eqnarray}
i\,\langle K^-(p_K)K^{*0}(p_{K^*},\varepsilon_{K^*})\left|
(V-A)_\mu\right|0\rangle &=& i\epsilon_{\mu\nu\alpha\beta}\,
\varepsilon_{K^*}^{*\nu}p_K^\alpha p_{K^*}^{\beta}
\,\frac{2\,V(q^2)}{m_K+m_{K^*}} \nonumber
\\
&+& %-i\bigg[ \varepsilon_{K^*}^*-\frac{\varepsilon_{K^*}^*\cdot
%q}{q^2}q\bigg]_\mu
\bigg( g_{\mu\nu}-\frac{q_\mu q_\nu}{q^2}\bigg)
\varepsilon_{K^*}^{*\nu}(m_K+m_{K^*})\,A_1(q^2)\nonumber
\\
&-& %+i\bigg[(p_K-p_{K^*})-\frac{(p_K-p_{K^*})\cdot
%q}{q^2}q\bigg]_\mu
\bigg( g_{\mu\nu}-\frac{q_\mu
q_\nu}{q^2}\bigg)(p_K-p_{K^*})^\nu\left(\varepsilon_{K^*}^*\cdot
q\right)\frac{A_2(q^2)}{m_K+m_{K^*}} \nonumber
\\
&+& \frac{2m_{K^*}}{q^2}\,q_\mu\left(\varepsilon_{K^*}^* \cdot
q\right) A_0(q^2)\,,
%\\
\label{eq:KKstar}
\end{eqnarray}
where the form factors $V$, $A_0$ and $A_1$, $A_2$ are induced, in
the $K^-K^{*0}$ rest frame, by the vector $V^i$~($1^-$), the
timelike component of the axial vector $A^{\mu=0}$~($0^-$) and the
spacelike component of the axial vector currents $A^i$~($1^+$),
respectively. It should be noted that, once factorized, the
angular momentum quantum number of the $K^-K^{*0}$, hence the
allowed resonances, can only be $J=0,~1$ to
match the current. Any $K^-K^{*0}$ or
intermediate resonance component with $J>1$ cannot be
accommodated~\cite{remark} within factorization approach.

Upon squaring the amplitude of Eq.~(\ref{eq:amplitude}), the decay
rate would involve interference between the four form factors of
$\langle K^-K^{*0}|V-A|0\rangle$. Simply knowing the $K^-K^{*0}$
mass spectrum is far from being sufficient to constrain all the
parameters. One thus needs further experimental inputs. Since we
have parameterized $\langle K^-K^{*0}|V-A|0\rangle$ in accordance
with the angular momentum configuration of the $K^-K^{*0}$, it is
straightforward to see what information can be gained by studying
the angular distributions.

Denoting $\mathbf{p}_h^*$ the three-momentum of meson $h$ in the
$K^-K^{*0}$ rest frame, and $\mathbf{p}_h$ in the $\overline B^0$
rest frame, the helicity angle $\theta_{KK}$ is defined as the
angle between $\mathbf{p}_{K^*}^*$ and
$-\mathbf{p}_D$~\cite{Drutskoy:2002ib}. The angular distributions
coming from %the spacelike component of the vector current
$V^i$~($1^-$) and %the timelike component of the axial current
$A^{\mu=0}$~($0^-$) are
\begin{eqnarray}
\frac{d\Gamma_V}{d\cos\theta_{KK}}&\propto&
\sin^2\theta_{KK}%=1-\cos^2\theta_{KK}
\,,\\
\frac{d\Gamma_{A_0}}{d\cos\theta_{KK}}&\propto&\mathrm{constant}\,,
\end{eqnarray}
respectively, where we have used $V$ and $A_0$ to denote the
corresponding contributions.
Likewise, %the spacelike component of the axial current
for $A^i$~($1^+$), we have
\begin{eqnarray}
%A_1:&&
\frac{d\Gamma_{A_1}}{d\cos\theta_{KK}} &\propto&
1+b\cos^2\theta_{KK}
%\int
%dM_{KK^*}\left|\mathbf{p}_{K^*}^*\right|\left|\mathbf{p}_D\right|
%\bigl(F_1^{BD}\bigr)^2%(m_{K}+m_{K^*})^2
%\bigl|A_1\bigr|^2 \mathbf{p}_D^{*2}\biggl(1+
%\frac{\mathbf{p}_{K^*}^{*2}} {m_{K^*}^2}\cos^2\theta_{KK}\biggr)
\,,
\label{eq:A1}\\
%
%A_2:&&
\frac{d\Gamma_{A_2}}{d\cos\theta_{KK}} &\propto&\cos^2\theta_{KK}
%\int
%dM_{KK^*}\left|\mathbf{p}_{K^*}^*\right|\left|\mathbf{p}_D\right|
%\bigl(F_1^{BD}\bigr)^2\bigl|A_2\bigr|^2
%\mathbf{p}_D^{*2}\biggl(\frac{4\,M_{KK^*}^2}{(m_{K}+m_{K^*})^4}
%\frac{|\mathbf{p}_{K^*}^*|^{\,4}}{m_{K^*}^2}
%\biggr)\cos^2\theta_{KK}
\,,
\label{eq:A2}\\
%A_1A_2:&&
\frac{d\Gamma_{A_1A^*_2}}{d\cos\theta_{KK}}
&\propto&\cos^2\theta_{KK}
%\int
%dM_{KK^*}\left|\mathbf{p}_{K^*}^*\right|\left|\mathbf{p}_D\right|
%%\bigl[F^{BD}\bigr]^2
%\bigl(F_1^{BD}\bigr)^2\mathrm{Re}\bigl[A_1A_2^*\bigr]
%\mathbf{p}_D^{*2}\biggl(\frac{4\,M_{KK^*}E^*_{K^*}
%\mathbf{p}_{K^*}^{*2}}{(m_{K}+m_{K^*})^2m_{K^*}^2}\,
%\biggr)\cos^2\theta_{KK}
\,,
\label{eq:A1A2}
%\nonumber\\
%&&
\end{eqnarray}
from $A_1$, $A_2$ and their interference, respectively.

%
%-----------------------------------------------------------------------
\begin{figure*}[tb]
\includegraphics[width=3.3in]{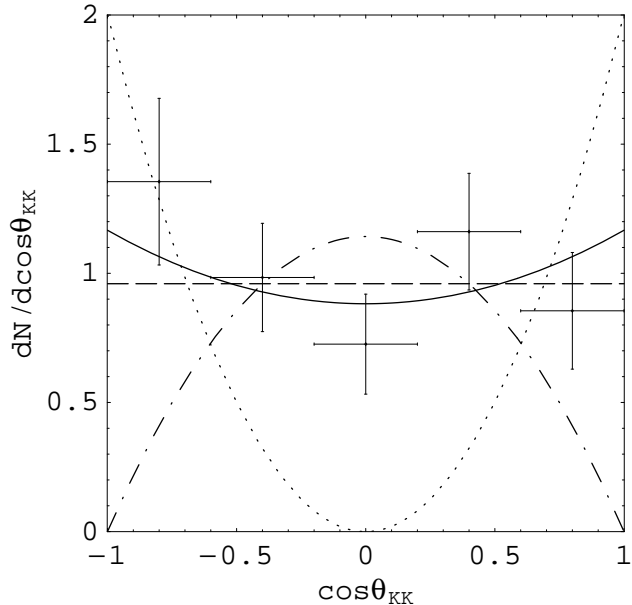}
\caption{\label{fig:angular_fit} Fit to angular distribution by
$a(1+b\cos^2\theta_{KK})$~(solid), a constant $c$~(dash),
$\xi\sin^2\theta_{KK}$~(dash-dot) and
$\zeta\cos^2\theta_{KK}$~(dot), with
$(a,b,\chi^2/\mathrm{n.d.f.})=(0.88,0.32,3.5/3)$,
$(c,\chi^2/\mathrm{n.d.f.})=(0.99,4.0/4)$,
$(\xi,\chi^2/\mathrm{n.d.f.})=(1.14,18.2/4)$ and
$(\zeta,\chi^2/\mathrm{n.d.f.})=(2.0,39.3/4)$, respectively. The
$\cos^2\theta_{KK}$ and $\sin^2\theta_{KK}$ distributions are
clearly disfavored by data. }
\end{figure*}
%------------------------------------------------------------------------
%

As shown in Fig.~\ref{fig:angular_fit}, the data points scatter
around the horizontal line $d$N$/d\cos\theta_{KK}\sim1$. One
concludes that the $V$ ($1^-$) and $A_2$ contributions are not
favored by present data
%, as they deviate from data at both
%$\cos\theta_{KK}\sim\pm1$ and $\cos\theta_{KK}\sim0$.
%So both terms involving $V$ and $A_2$
and can be safely dropped at this stage. On the other hand,
angular analysis of subsequent $K^*\to K\pi$ decay suggests
$K^-K^{*0}$ is dominantly $1^+$~\cite{Drutskoy:2002ib}, therefore
preferring $A_1$ over $A_0$.

\begin{figure*}[tb]
\includegraphics[width=3.1in]{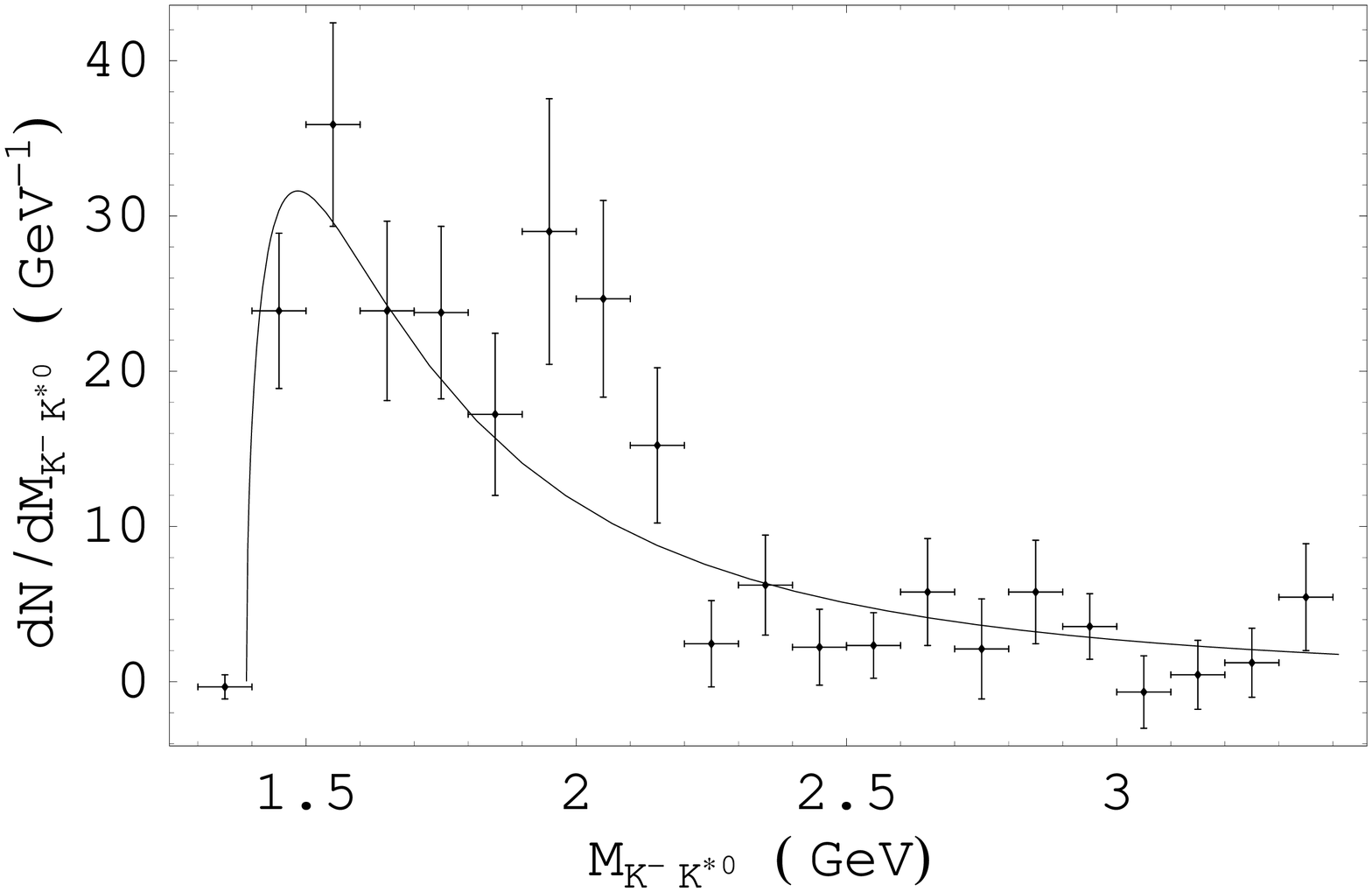}
\caption{\label{fig:dNdM02} The observed~\cite{Drutskoy:2002ib}
$K^-K^{*0}$ mass spectrum combining $D^+K^-K^{*0}$,
$D^{*+}K^-K^{*0}$, $D^0K^-K^{*0}$ and $D^{*0}K^-K^{*0}$ modes. The
experimental fit in \cite{Drutskoy:2002ib} assumes only the
$a_1(1260)$ resonance. The $\chi^2/$n.d.f. is $\sim24/18$ and the
obtained nominal width of $a_1(1260)$ is $380$~MeV.}
\end{figure*}

It is interesting to understand how $A_1$, with
$d\Gamma_{A_1}/d\cos\theta_{KK}\propto 1+b\cos^2\theta_{KK}$, can
describe the data in Fig.~\ref{fig:angular_fit}, which seems quite
consistent with a constant distribution. From
\begin{eqnarray}
\frac{d\Gamma_\mathrm{tot}}{d\cos\theta_{KK}}&\thickapprox&
\frac{d\Gamma_{A_1}}{d\cos\theta_{KK}} \nonumber\\
&\propto& \int
dM_{KK^*}\left|\mathbf{p}_{K^*}^*\right|\left|\mathbf{p}_D\right|
\bigl(F_1^{BD}\bigr)^2%(m_{K}+m_{K^*})^2
\bigl|A_1\bigr|^2 \mathbf{p}_D^{*2}\biggl(1+
\frac{\mathbf{p}_{K^*}^{*2}}
{m_{K^*}^2}\cos^2\theta_{KK}\biggr)\label{eq:b-dep}
\\
&\propto& a(1+b\cos^2\theta_{KK})\,, \label{eq:b}
\end{eqnarray}
where $M_{KK^*}^2\equiv q^2$. %the invariant mass of the $K^-K^{*0}$,
We find that $b$ in Eq.~(\ref{eq:b}) is {\it small} due to the
factor $\mathbf{p}_{K^*}^{*2}$ in Eq.~(\ref{eq:b-dep}) as a
consequence of threshold peaking, resulting in a flattened
parabola which mimics a constant.
We reproduce the experimental data in Fig.~\ref{fig:dNdM02},
where one can see a clear threshold peak while large $t \equiv
M_{KK^*}^2$ is suppressed. The major part of the rate comes from the
low $M_{KK^*}$ region, resulting in a suppressed
$\mathbf{p}_{K^*}^{*2}$ in $KK^*$ frame. For higher $M_{KK^*}$
where $\mathbf{p}_{K^*}^{*2}$ is large, the contribution to $b$ is
suppressed by the tail of the $K^-K^{*0}$ mass spectrum.

%
%-----------------------------------------------------------------------
%------------------------------------------------------------------------
%

It should be emphasized that, although the present data disfavor
any significant contributions from $V$, $A_2$ and $A_0$, it is
possible that these contributions, however small, may show up in
the future when one has sufficient amount of data. It is clear,
however, that one can always use angular analysis to project out
each form factor. %that have so far been neglected.
As this is not yet the case, for this work, we concentrate only on
the dominant $A_1(q^2)$ term and neglect terms involving other
form factors. The $\overline B^0\to D^+K^-K^{*0}$ decay amplitude
can now be simplified to
\begin{equation}
{\mathcal A}(D^+K^-K^{*0})=-i\,\frac{G_F}{\sqrt2}V_{cb}V_{ud}^*
 \,a_1 F^{BD}_1(q^2)\,A_1(q^2)(m_K+m_{K^*})\,2p_D\cdot\bigg(
\varepsilon_{K^*}^*-\frac{\varepsilon_{K^*}^*\cdot
q}{q^2}\,q\bigg)\,. \label{eq:DecayAmplitude}
\end{equation}
%
%
%We have reproduced the experimental data in Fig.~\ref{fig:dNdM02},
%where the pattern of a clear threshold peak versus large $t \equiv
%M_{KK^*}^2$ suppression can be seen.
 As shown in Fig.~\ref{fig:dNdM02} the data
generally follows the experimental fit that assumes only the
$a_1(1260)$ resonance, which we have now generalized into the
$A_1(q^2)$ form factor. Note that the second peak at $M_{KK^*}\sim
2$~GeV in the data has large statistical error, and is dubious for
lack of known resonances. In any case, it would require more data to
clarify.
% Furthermore, since the data is given by combining
%$D^+K^-K^{*0}$, $D^{*+}K^-K^{*0}$, $D^0K^-K^{*0}$ and
%$D^{*0}K^-K^{*0}$ decays, the origin of this peak is unclear, even
%if it exists.

$K^-K^{*0}$ can be produced via both resonant and non-resonant
contributions. We parameterize the $A_1$ form factor by including
a single $a_1(1260)$ resonance to account for the resonant
contribution. For non-resonant contributions, we take cue from
perturbative QCD~(PQCD)~\cite{Brodsky:1974vy}, which states that
at least one {\it hard} gluon is needed to redistribute the large
momentum transfer $t$ in $K^-K^{*0}$. We then follow
\cite{Chua:2001vh} and expand in $1/t^n$ for the non-resonant
part, with $n$ starting at $1$ to reflect this hard gluon
exchange. For simplicity, and since data is still scarce at
present, we take just two terms~($n=1$, $2$). We further impose a
$180^\circ$ difference between the phases of the complex expansion
coefficients. This is to compensate for the (possibly) artificial
rise at low $t$ from the leading $1/t$ term. Such alternative
signs are seen in the fits of nucleon EM form
factors~\cite{Chua:2001vh}. The non-resonant phases are taken as
constant.
The $A_1$ form factor is then parameterized as
\begin{equation}
A_1(t)=\frac{g_s\,m_{a_1}f_{a_1}}{t-m_{a_1}^2+i\,m_{a_1}\Gamma(t)}
+e^{i\phi}\left(\frac{x_1}{t}-\frac{x_2}{t^2}\right)
\left[\ln\left(\frac{t}{0.3^2}\right)\right]^{-1},
\label{eq:parametrization}
\end{equation}
where $m_{a_1}$, $f_{a_1}$ are respectively the mass and decay
constant of $a_1(1260)$, $g_s$ the $a_1(1260)\to KK^*$ coupling
constant, and $x_{1,2}$ and $\phi$ are the strengths and phase of
the non-resonant terms.

The $a_1(1260)$ is broad and the width is not yet well known,
and could range anywhere between $250$ to
$600$~MeV~\cite{Hagiwara:pw}. It is not appropriate to treat the
total width as constant, which is only valid in narrow width
approximation. We use~\cite{Drutskoy:2002ib},
\begin{equation}
\Gamma(t)=\Gamma_{a_1}\,\frac{\rho_{\rho\pi}(t)}{\rho_{\rho\pi}(m_{a_1}^2)}\,,
\label{eq:Gamma}
\end{equation}
where the constant $\Gamma_{a_1}$ is the nominal width of
$a_1(1260)$, and
\begin{equation}
\rho_{\rho\pi} \equiv
\left(1+\frac{\mathbf{p_\rho^*}^2}{3\,m_\rho^2}\right)|\mathbf{p_\rho^*}|\,,
\end{equation}
where, in addition to the phase space factor $|\mathbf{p_\rho^*}|$
similar to what is used in \cite{Drutskoy:2002ib}, we also take
into account the amplitude squared of
$a_1(1260)\to(\rho\pi)_{s-\mathrm{wave}}$. The total width of
$a_1(1260)$ is approximated by the
$a_1(1260)\to(\rho\pi)_{s-\mathrm{wave}}$ partial width.

The branching fraction of $a_1^-(1260)\to K^-K^{*0}$ can be
extracted by taking the ratio of the areas under the spectral
functions of $a_1^-(1260)\to$ all (approximated by
$a_1\to(\rho\pi)_{s-\mathrm{wave}}$) and $a_1^-(1260)\to
K^-K^{*0}$, respectively:
\begin{eqnarray}
\Pi_{\rho\pi}&\equiv&\frac{m_{a_1}\Gamma_{a_1}}{\pi}
\frac{\Gamma(t)}{(t-m_{a_1}^2)^2+m_{a_1}^2\Gamma(t)^2}\,,\\
\Pi_{KK^*}&\equiv&\frac{m_{a_1}\Gamma_{a_1}}{\pi}
\frac{\Gamma_{KK^*}(t)}{(t-m_{a_1}^2)^2+m_{a_1}^2\Gamma(t)^2}\,,
\end{eqnarray}
where the partial width of $a_1(1260)\to K^-K^{*0}$ is given by
\begin{equation}
\Gamma_{KK^*}(t)\equiv\left[\frac{g_s^2(m_K+m_{K^*})^2}{8\pi}\frac{\rho_{\rho\pi}(m_{a_1}^2)}{m_{a_1}^2}\right]
\frac{\rho_{KK^*}(t)}{\rho_{\rho\pi}(m_{a_1}^2)}\,,
\label{eq:GammaKKstar}
\end{equation}
and
\[
\rho_{KK^*} \equiv
\left(1+\frac{\mathbf{p_{K^*}^*}^2}{3\,m_{K^*}^2}\right)|\mathbf{p_{K^*}^*}|\,.
\]
Comparing to the definition of $\Gamma(t)$, the constant inside
the brackets of Eq.~(\ref{eq:GammaKKstar}) plays the same role as
that of $\Gamma_{a_1}$ in Eq.~(\ref{eq:Gamma}), i.e. the nominal
value of the partial width of $a_1^-(1260)\to K^-K^{*0}$.
In the narrow width limit, $\Gamma_{a_1}\to0$, we recover
\[
\frac{1}{(t-m_{a_1}^2)^2+m_{a_1}^2\Gamma(t)^2}\thickapprox\frac{\pi}{m_{a_1}\Gamma_{a_1}}\delta(t-m_{a_1}^2)\,,
\]
such that
\begin{eqnarray}
\Pi_{\rho\pi}&\thickapprox&\Gamma(m_{a_1}^2)\delta(t-m_{a_1}^2)=\Gamma_{a_1}\delta(t-m_{a_1}^2)\,,\\
\Pi_{KK^*}&\thickapprox&\Gamma_{KK^*}(m_{a_1}^2)\delta(t-m_{a_1}^2)\,.
\end{eqnarray}
for $\Gamma_{a_1}\to0$. Since in reality $m_{KK^*}>m_{a_1}$ the
``branching fraction" $R$ of $a_1^-(1260)\to K^-K^{*0}$ is then
defined by
\begin{equation}
R\equiv\frac{\int^\infty_{(m_K+m_{K^*})^2} \Pi_{KK^*}(t)dt}
{\int^\infty_{(m_K+m_{K^*})^2}
\Pi_{KK^*}(t)dt+\int^\infty_{(m_\rho+m_\pi)^2}
\Pi_{\rho\pi}(t)dt}\,. \label{eq:a1KKstar}
\end{equation}

Our method for extracting the $a_1^-(1260)\to K^-K^{*0}$ rate is
similar to the one used by the CLEO Collaboration in
Ref.~\cite{Asner:1999kj}. In the fit to
$\tau^-\to\nu_\tau\pi^-\pi^0\pi^0$ which is dominated by
$\tau^-\to\nu_\tau a_1^-(1260)$, the $a_1^-(1260)\to K^-K^{*0}$
``rate"~($R_{\rm CLEO}\sim3\%$) was extracted by taking into
account $a_1^-(1260)\to K^-K^{*0}$ and other processes that may
contribute to the $a_1(1260)$ Breit-Wigner width. This is in
contrast with the way the result was obtained in
Ref.~\cite{Drutskoy:2002ib}, where the $a_1(1260)$ was thought to
account for the whole spectrum, and the corresponding factor
$R_{\rm Belle}$ is obtained by using $R_{\rm Belle}\equiv{\mathcal
B}(\overline B^0 \to D^+ K^-K^{*0})/{\mathcal B}(\overline B^0 \to
D^+ a_1^-(1260))\sim15\%$. Only in the narrow width and
$m_{a_1}>m_{KK^*}$ limit that $R_{\rm Belle}$ reduces to $R$
defined in Eq.~(\ref{eq:a1KKstar}).

%%%%%%%%%%%%%%%%%%%%%%
\section{Results}
\label{sec:Results}
%%%%%%%%%%%%%%%%%%%%%%

Throughout this paper, we use $a_1=0.935$, $m_{a_1}=1230$~MeV,
$f_{a_1}=229$~MeV and the CKM matrix elements $V_{ud}=0.975$,
$V_{cb}=0.039$ as in Ref.~\cite{Chua:2002pi}.

The mass spectra plotted in Fig.~\ref{fig:dNdM02} is taken from
Ref.~\cite{Drutskoy:2002ib}, which combines $D^+K^-K^{*0}$,
$D^{*+}K^-K^{*0}$, $D^0K^-K^{*0}$ and $D^{*0}K^-K^{*0}$ modes. Our
model parametrization in previous section, however, is constructed
only for $D^+K^-K^{*0}$. Since our present purpose is only to
demonstrate the feasibility of extracting the $A_1$ form factor
from $B$ decay data, we shall make our own {\it mock data} on
which to practice the extraction.

Starting with Fig.~\ref{fig:dNdM02}, which is based on
$29.4$~fb$^{-1}$ of data, we first transform from
$d\mathrm{N}/dM_{KK^*}$ into $d\mathcal{B}/dM_{KK^*}$, normalized
by the measured
$\mathcal{B}(D^+K^-K^{*0})=(8.8\pm1.1(\mathrm{stat.})
\pm1.5(\mathrm{syst.})) \times10^{-4}$~\cite{Drutskoy:2002ib}.
Since the KEKB accelerator has already accumulated over $150$
million $B\overline B$ pairs~$(\sim140~\mathrm{fb}^{-1})$ by
summer 2003, we take our mock data to be five times the original
one~$(\sim29.4~\mathrm{fb}^{-1})$, i.e.
$\sim150~\mathrm{fb}^{-1}$, but keeping the central value for
$\mathcal{B}(D^+K^-K^{*0})$ unchanged. Next, we refine the
resolution in $M_{KK^*}$ by rebinning two bins into three, again
keeping the same central value and maintaining consistent
statistical error.

%
%-----------------------------------------------------------------------
\begin{figure*}[b!]
\includegraphics[width=3.3in]{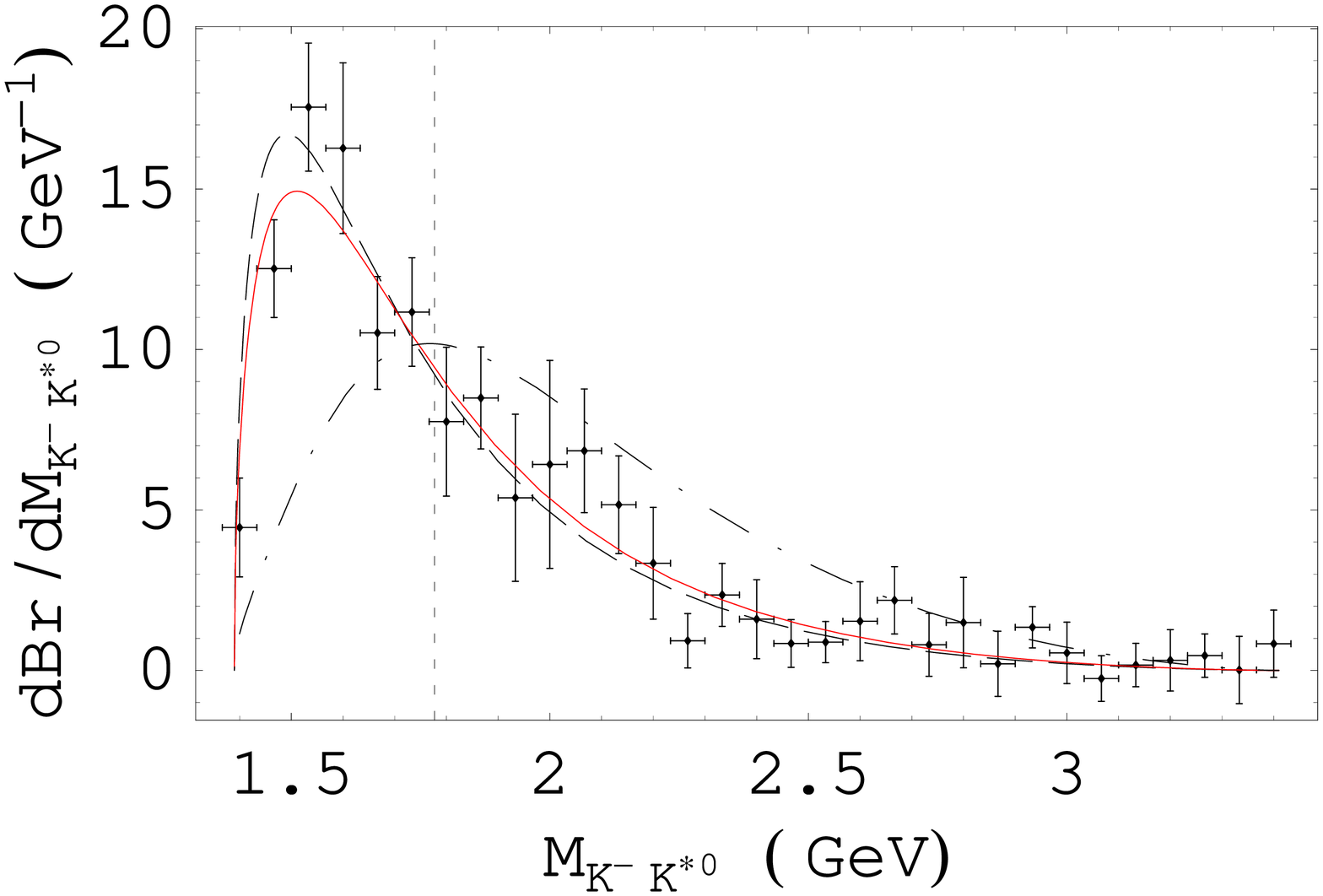}
\caption{\label{fig:dBrdMrebin3Dplus} $d\mathcal{B}/dM_{KK^*}$~(in
units of $10^{-4}$) mock data for $\overline B^0\to D^+K^-K^{*0}$
assuming five times ($\sim150$~fb$^{-1}$) the original
data~\cite{Drutskoy:2002ib} of $29.4$~fb$^{-1}$. The solid curve
corresponds to Eq.~(\ref{eq:bestfit}). The dash curve is given by
$g_s=1.4$, $\Gamma_{a_1}=430$~MeV, $x_{1(2)}=8(1)$~GeV$^{2(4)}$
and $\phi=-40^\circ$. The dot-dash curve is given by
$g_s=0.89$, $\Gamma_{a_1}=350$~MeV,
$x_{1(2)}=23.2(38.2)$~GeV$^{2(4)}$ and $\phi=201^\circ$.
The vertical dot line indicates the $\tau$ mass.}
\end{figure*}
%------------------------------------------------------------------------

The second peak at $M_{KK^*}\sim2$~GeV remains after rebinning,
but no known resonances can account for it. We therefore {\it remove it
by hand}. In order to do so, we first assume the data has a shape
that roughly follows the curve given by some chosen set of parameters:
$g_s=1.4$, $\Gamma_{a_1}=430$~MeV, $x_1=8$~GeV$^2$, $x_2=1$~GeV$^4$,
and $\phi=-40^\circ$, which give
$\mathcal{B}(D^+K^-K^{*0})=8.4\times10^{-4}$ and
$\chi^2/\mathrm{n.d.f.}=22.6/26$. The branching fraction $R$ of
$a_1^-(1260)\to K^-K^{*0}$ given by Eq.~(\ref{eq:a1KKstar}) is
$4.3\%$, which is small compared to the value~($8\%-15\%$) of
Ref.~\cite{Drutskoy:2002ib}. With this curve as guide,
we remove the second peak, but respect the statistical error in the
corresponding $M_{KK^*}$ range.

Having generated the mock data, we then redo the fitting. The best
fit with $\chi^2/$n.d.f.$=19.2/26$ is obtained with the following:
\begin{equation}
\begin{array}{ccccc}
g_s=2.41, & \Gamma_{a_1}=678~\mathrm{MeV}, &
x_1=8.32~\mathrm{GeV}^2, & x_2=3.07~\mathrm{GeV}^4, &
\phi=-18.7^\circ,
\end{array}
\label{eq:bestfit}
\end{equation}
which give $\mathcal{B}(D^+K^-K^{*0})=8.33\times10^{-4}$, and a
larger branching fraction $R=8.18\%$ due to a larger strong
coupling $g_s$. The fit result is shown in
Fig.~\ref{fig:dBrdMrebin3Dplus}. The dominant contribution is from
the non-resonant term, which gives $\sim54\%$ of the rate. The
$a_1(1260)$ resonance contributes only $\sim9\%$ of the rate by
itself, with the remaining $\sim36\%$ arising from interference
with the non-resonant term. The interference is constructive
because the non-resonant phase $\phi=-18.7^\circ$ lies in the same
quadrant as the resonance phase
$\tan^{-1}[-m_{a_1}\Gamma(t)/(t-m_{a_1}^2)]$, which varies between
$-90^\circ$ to $0^\circ$ for $t\gtrsim m_{a_1}^2$.
%as can be seen from Eq.~(\ref{eq:parametrization}),

%
%-----------------------------------------------------------------------
\begin{figure*}[b!]
\includegraphics[width=3.3in]{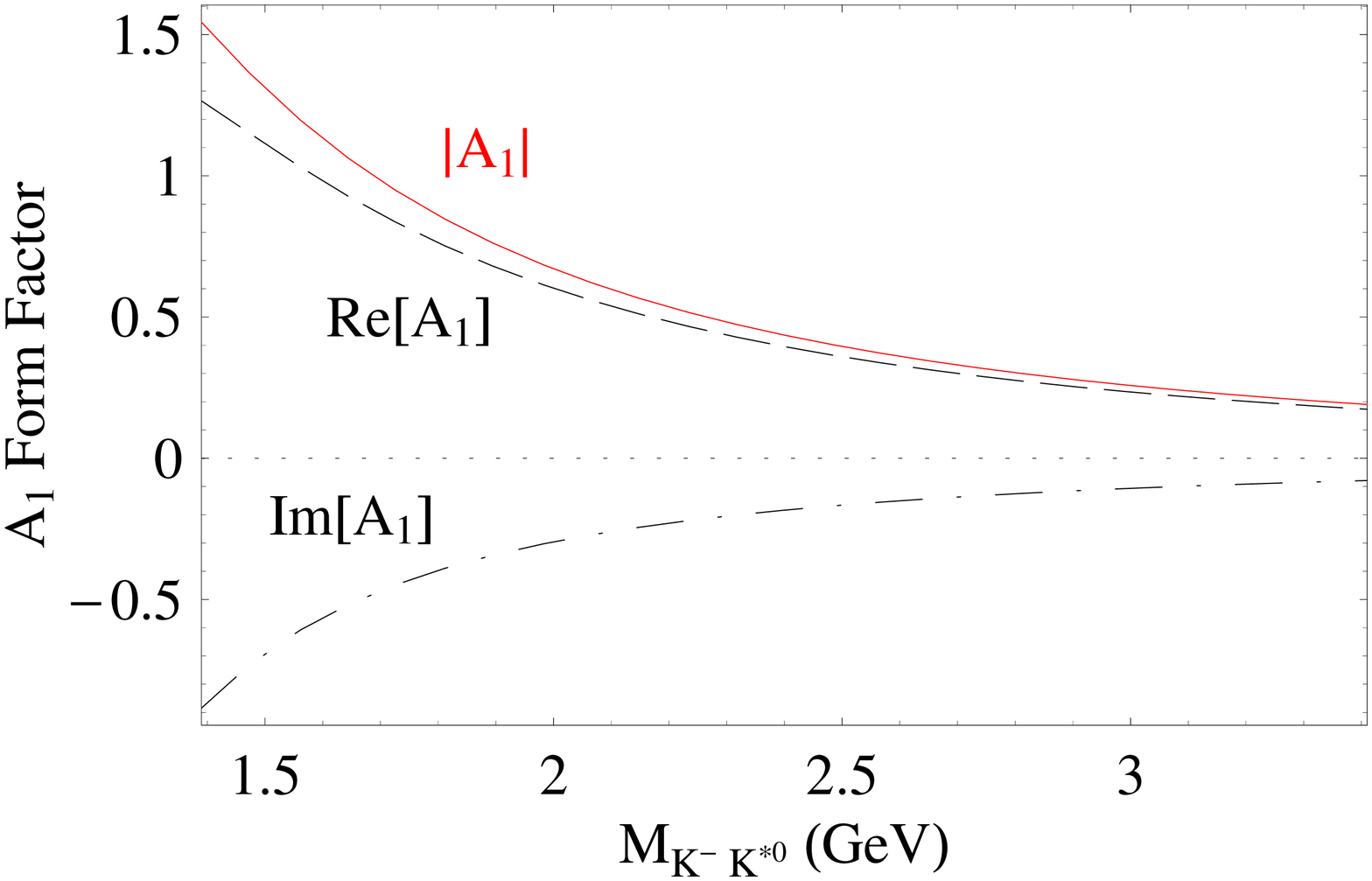}
\caption{\label{fig:A1} Plot of $|A_1|$~(solid),
$\mathrm{Re}[A_1]$~(dash) and $\mathrm{Im}[A_1]$~(dot-dash) from
Eq~(\ref{eq:bestfit}).}
\end{figure*}
%------------------------------------------------------------------------

In Fig.~\ref{fig:A1} we show the $A_1$ form factor with the
parameters of Eq.~(\ref{eq:bestfit}). The relative sign between
the real and imaginary parts of the form factor is mainly due to
the phase $\phi=-18.7^\circ$, which is coherent with the
$a_1(1260)$ resonance phase, and gives a negative imaginary part.
The $x_1$ term contributes $\sim131\%$ to the non-resonant
contribution and dominates $A_1$, compared to only $\sim2\%$ from
$x_2$, where we have built in destructive interference between
$x_1$ and $x_2$ parts.

It is useful to investigate the $\chi^2$ behavior in the vicinity
of the best fit values of Eq.~(\ref{eq:bestfit}). We find that,
within $\Delta\chi^2=1$, the $\chi^2$ depends on $g_s$ and
$\Gamma_{a_1}$ roughly via the ratio $g_s/\Gamma_{a_1}$, hence
many ``solutions" exist with differing $a_1$ width. This is
because the rate is dominated by the near-threshold contribution
where the $t-m_{a_1}^2$ factor in Eq.~(\ref{eq:parametrization})
is small compared to $m_{a_1}\Gamma(t)$, hence resulting in a
resonance term that is roughly characterized by
$g_s/\Gamma_{a_1}$. Since this is inherent in our form factor
model, one cannot determine $g_s$ and $\Gamma_{a_1}$ separately
even when the independent $\overline B^0\to D^+K^-K^{*0}$ spectrum
has become available. One therefore needs independent input on
$a_1(1260)$ resonance, which we discuss in the next section.

%%%%%%%%%%%%%%%%%%%%%%
\section{Discussion and Conclusion}
\label{sec:discussion}
%%%%%%%%%%%%%%%%%%%%%%

We see that independent input on $a_1(1260)$ width is needed to
extract the $K$-$K^*$ axial $A_1$ form factor.
To clarify this point further, we note that the $\tau\to\nu_\tau
K^-K^{*0}$ process is very similar to $\overline B^0\to
D^+K^-K^{*0}$, with $a_1V_{cb}\langle D^+|(V-A)^\mu|\overline
B^0\rangle$ in Eq.~(\ref{eq:amplitude}) replaced by $\bar
u_{\nu_\tau}\gamma^\mu(1-\gamma_5)u_\tau$.
%It is of interest to calculate the rate of
%$\tau\to\nu_\tau K^-K^{*0}$ as a check for our $A_1$ form factor.
Factorization in this case is basically exact, and $K^-K^{*0}$
production comes only from the axial current. By using the best
fit values of Eq.~(\ref{eq:bestfit}), we obtain
$\mathcal{B}(\tau\to\nu_\tau
K^-K^{*0})=5.3\times10^{-3}\sim0.5\%$, similar to Ref.
\cite{Drutskoy:2002ib}, and is more than twice the measured rate
of $(2.1\pm0.4) \times 10^{-3}$~\cite{Hagiwara:pw}.

Our larger predicted $\tau\to\nu_\tau K^-K^{*0}$ rate can be
understood as follows. Although both $\mathcal{B}(\overline B^0\to
D^+ K^-K^{*0})$ and $\mathcal{B}(\tau\to\nu_\tau K^-K^{*0})$ are
proportional to $|A_1|^2$, $\tau\to\nu_\tau K^-K^{*0}$ receives
contributions only from the region of $M_{KK^*} < m_\tau$. Our
mock data implies $\sim60\%$ of the $\overline B^0\to D^+
K^-K^{*0}$ rate comes from $M_{KK^*} < m_\tau$. The dominance of
near-threshold contribution suggested by $\overline B^0\to
D^+K^-K^{*0}$ data translates to a large $\tau\to\nu_\tau
K^-K^{*0}$. Thus, to reduce $\mathcal{B}(\tau\to\nu_\tau
K^-K^{*0})$ from $\sim0.5\%$ down to $\sim0.2\%$, one needs to
reduce the relative contribution coming from near-threshold. We
illustrate this by the dot-dash line in
Fig.~\ref{fig:dBrdMrebin3Dplus}, where only $\sim32\%$ of the rate
comes from $M_{KK^*}\leq m_\tau$ and the peak is shifted towards
higher $t$. This would have given the correct rates
$\mathcal{B}(\tau\to\nu_\tau K^-K^{*0})\sim2.1\times10^{-3}$ and
$\mathcal{B}(\overline B^0\to D^+ K^-K^{*0})\sim8.8\times10^{-4}$,
but seems to disagree with the mock $\overline B^0\to D^+
K^-K^{*0}$ spectrum.

Recall, however, the comparison between the predicted spectrum of
$D^+ K^-K^0$ with that of $D^0 K^-K^0$ in Ref.~\cite{Chua:2002pi}. Due
to the additional $B\to KK$ transition mechanism that is
asymptotically $1/t^2$, the $D^0 K^-K^0$ spectrum has a peak
closer to the threshold~($\sim250$~MeV above threshold) than that
of $D^+ K^-K^0$~($\sim600$~MeV above threshold).
Since the mock data used in the present work is generated from
published data that {\it combines} those of $D^+ K^-K^{*0}$,
$D^{*+} K^-K^{*0}$, $D^0 K^-K^{*0}$ and $D^{*0} K^-K^{*0}$, it is
conceivable that the real data for $\overline B^0\to D^+
K^-K^{*0}$ could be somewhat closer to the dot-dash line in
Fig.~\ref{fig:dBrdMrebin3Dplus}, which peaks at $\sim400$~MeV
above threshold. In other words, the peak near the threshold of
Fig.~\ref{fig:dNdM02}~(or Fig.~\ref{fig:dBrdMrebin3Dplus}) could
be due to the $D^{(*)0}K^-K^{*0}$ modes, whereas the apparent peak
at $\sim2$~GeV in Fig.~\ref{fig:dNdM02} (which we removed in
Fig~\ref{fig:dBrdMrebin3Dplus})
%, although its existence is still hard to be justified by
%resonances, with its statistical errors are still large,
might be some hint of the behavior of the $D^{(*)+}K^-K^{*0}$
modes. The situation can be clarified by separating $D^+K^-K^{*0}$
and $D^0K^-K^{*0}$ with more data.
We note further from Fig.~\ref{fig:A1} that
$|A^{KK^*}_1(q^2= (1.5\ {\rm GeV})^2)|\simeq 1.3$ is larger than
$|F^{KK}_1(q^2= (1.5\ {\rm GeV})^2)|\simeq 0.7$
as fitted from kaon EM data Ref.~\cite{Chua:2002pi}.
This gives further support of our conjecture
that contamination of $B^-\to D^{(*)0}K^-K^{*0}$ in the mock
data could be responsible for the enhancement of our
$\tau\to\nu_\tau K^-K^{*0}$ rate.

From the point of view of extracting the timelike $K^-K^{*0}$
axial $A_1$ form factor, $\tau\to\nu_\tau K^-K^{*0}$ is better
because the production of $KK^*$ is purely proportional to $A_1$.
However, it has limited range in momentum transfer. In contrast,
the $\overline B^0\to D^+ K^-K^{*0}$ mode allows us to probe
higher $t$, but one has the associated factor~$a_1F^{BD}_1$~(see
Eq.~(\ref{eq:DecayAmplitude})), which means one is further subject
to possible corrections to factorization. We have assumed certain
values for $a_1F^{BD}_1$ to extract $A_1$.  A more appropriate way
to proceed would be to treat $A_1$ and $a_1F^{BD}_1(q^2)$ on equal
footing, both as suitably parameterized functions to be determined
from data. To be able to do so, one needs further independent
data, such as an analysis of $\overline B\to D \rho\pi$ decay
aimed at extracting $\overline B\to D a_1(1260)$, which should be
performed in a similar way to our $\overline B^0\to D^+ K^-K^{*0}$
study, including angular analysis and proper parametrization of
$\pi$-$\rho$ axial form factor that respects PQCD for large
$M_{\rho\pi}$. There is, of course, the need for concurrent check
on validity of factorization. This may then involve extraction of
$F^{BD}_1(q^2)$ from $\overline B^0\to D^+\ell^-\nu$.

Our result reveals that the contribution from the $a_1(1260)$
resonance~($\sim 9\%$) is much smaller than that from the
non-resonant part~($\sim54\%$). While this underlines
the importance of non-resonant part, the smallness of the
$a_1(1260)$ resonance contribution is curious. Recall that in the
$\overline B^0\to D^+K^-K^{0}$ decay we have $\rho$-resonance
contribution up to 40\%~\cite{Chua:2002pi}. It should be pointed
out that, through SU(3) relations, the $\rho$ resonance
contribution in $F^{KK}_1$ is highly constrained by the well
measured $\phi$ resonance contribution in kaon EM
data~\cite{Chua:2002pi}. It is not clear if the smallness of the
$a_1$ resonance contribution is an artifact of having only one
resonant term. On the other hand, the smallness of $R\sim R_{\rm
CLEO}<10\%$ seems to support a suppressed $a_1(1260)$-resonance
contribution. It is interesting to see that, if we follow
Ref.~\cite{Drutskoy:2002ib} to take the ratio between the rate
contributed from $a_1(1260)$ alone~($9\%$ of the total) and the
rate of $\overline B^0\to D^+a_1^-(1260)$ from Ref.~\cite{Hagiwara:pw},
we obtain $R\sim1.3\%$, which is about $9\%$ of the Belle result
$R_{\rm Belle}\sim15\%$. The discrepancy could be due to the
breakdown of the narrow width limit and the physical fact of
$m_{KK^*}<m_{a_1}$, which are essential for $R_{ \rm Belle}$ to be
identical with $R$ defined in Eq.~(\ref{eq:a1KKstar}).

The case of $\overline B {}^0\to D^+\rho^0 \pi^-$ is somewhat
better in the sense that $\rho\pi$ is dominated by the $a_1$
resonance. However, there are many resonances that
decay to the $\rho\pi$ final state. One can extract $\overline
B\to D a_1(1260)$ rate by suitably considering these resonant and
non-resonant contributions.
The branching fractions of ${\mathcal B}(\overline B^0\to
D^+a_1^-(1260))=(0.60\pm0.22\pm0.24)\%$ and ${\mathcal B}(B^-\to
D^0a_1^-(1260))=(0.45\pm0.19\pm0.31)\%$ were obtained by an
analysis of $\overline B\to D\pi\pi\pi$ decays based on $\sim
200$~pb$^{-1}$ of data~\cite{Bortoletto:kz}. The large systematic
error is dominated by uncertainties in fitting
and are estimated by considering alternate backgrounds.
The mass and the width of $a_1(1260)$ were taken as {\it a priori} in
the extraction, which may no longer be appropriate when the data
has improved by more than three orders of magnitude
by the B factories.
To reduce the effect caused by uncertain properties
such as the width and sub-decay modes of $a_1(1260)$,
an amplitude that takes into account as complete substructures of
$a_1(1260)$ as possible is needed.
Furthermore, as we have seen from the fit to the
$\overline B^0\to D^+K^-K^{*0}$ data, a non-resonant part that
respects PQCD at higher $t$ may contribute significantly to the results,
which must also be taken into account to complete
the construction of the amplitude.
Of course, the cost would be to boost the number of fit parameters,
but fortunately one now has enormous amount of data.

As noted in the end of the previous section, the $a_1(1260)$
contribution depends on $g_s$ and $\Gamma_{a_1}$ roughly through
the ratio $g_s/\Gamma_{a_1}$. Thus, $\overline B^0\to
D^+K^-K^{*0}$ mode is not so powerful for determining the
resonance parameters. One needs independent determination of these
parameters, especially $\Gamma_{a_1}$ from other sources.
Therefore, it appears that, to extract the axial $K$-$K^*$ form
factor $A_1$ and understand properly the $a_1(1260)$ component,
one needs to perform a combined analysis of $\overline B^0\to
D^+K^-K^{*0}$, $D^+ \rho^0\pi^-$, $D^+\ell^-\nu$, and
$\tau\to\nu_\tau K^-K^{*0}$.

%%%%%%%%%%%%%%%%%%%%%%%%%%%%%%%%%%%
%\section{Conclusion}
%\label{sec:Conclusion}
%%%%%%%%%%%%%%%%%%%%%%%%%%%%%%%%%%%

In conclusion, based on what is revealed by 29.4 fb$^{-1}$ data
from Belle, we have illustrated how to extract the timelike
$K^-$-$K^{*0}$ axial form factor $A_1(M_{KK^*}^2)$ from $\overline
B^0\to D^+K^-K^{*0}$ data. The method can be easily generalized to
other form factors by angular analysis when more data becomes
available.
With $>$ 300 fb$^{-1}$ data already accumulated by KEKB and PEP-II,
if the case for factorization in such three-body $B$ decays is
further strengthened, the B Factories may be promising for the
study of hadronic form factors that may be otherwise inaccessible.
Hadronic form factors have played an instrumental role in the
formulation of nuclear and particle physics, and further insight
may perhaps be gained from quantities such as the $K$-$K^*$ axial
form factor. Our study already indicates that the ron-resonant
contribution, needed to account for large $t$ behavior,
is rather significant.
The current discussion also illustrates how nucleon
axial form factors can be extracted from $B^0\to D^{(*)-}p\bar n$
decay once the spectrum becomes available.

\begin{acknowledgements}
We thank A. Drutskoy, H.-C. Huang and H.-n. Li for discussions.
This work is supported in part by the National Science Council of
R.O.C. under Grants NSC-92-2112-M-002-024 and
NSC-92-2811-M-001-054, the MOE CosPA Project, and the BCP Topical
Program of NCTS.
\end{acknowledgements}

%%%%%%%%%%%%%%%%%%%%%%%%%%%


\begin{thebibliography}{99}
\bibitem{remark}
If $J>1$ components for $K^-K^0$ is extracted in the future, one
would have to go beyond factorization to account for it.
\end{thebibliography}
\end{document}